# An Adaptive Feature Extraction Algorithm for Classification of Seismocardiographic Signals

Amirtaha Taebi*, *Student Member, IEEE*, Brian E. Solar, and Hansen A. Mansy, *Member, IEEE*
Biomedical Acoustics Research Laboratory, University of Central Florida, Orlando, FL 32816, USA
{taebi@knights., bsolar@knights., hansen.mansy@} ucf.edu

*Abstract*—This paper proposes a novel adaptive feature extraction algorithm for seismocardiographic (SCG) signals. The proposed algorithm divides the SCG signal into a number of bins, where the length of each bin is determined based on the signal change within that bin. For example, when the signal variation is steeper, the bins are shorter and vice versa. The proposed algorithm was used to extract features of the SCG signals recorded from 7 healthy individuals (Age: 29.4±4.5 years) during different lung volume phases. The output of the feature extraction algorithm was fed into a support vector machines classifier to classify SCG events into two classes of high and low lung volume (HLV and LLV). The classification results were compared with currently available non-adaptive feature extraction methods for different number of bins. Results showed that the proposed algorithm led to a classification accuracy of ~90%. The proposed algorithm outperformed the non-adaptive algorithm, especially as the number of bins was reduced. For example, for 16 bins, $F_1$ score for the adaptive and non-adaptive methods were 0.91±0.05 and 0.63±0.08, respectively.

*Keywords—Seismocardiography (SCG), adaptive feature extraction, classification, cardiorespiratory, support vector machine (SVM).*

## I. Introduction

Vibrations created by cardiac activities (such as valve opening and closure, blood flow momentum changes, and myocardial contraction) can be measured noninvasively at the chest surface. Seismocardiography (SCG) is a technique that measures these vibrations using, for example, an accelerometer [1]–[3]. SCG might provide information that complements other cardiac monitoring methods such as echocardiography and electrocardiography [4]–[7]. Several algorithms and methods have been used to study SCG features with different levels of success. Understanding different characteristics of SCG signals may lead to a better understanding of heart function, as well as a more successful characterization of these signals. Better characterization and classification of SCG signals in health and diseased can provide possible methods for diagnosing and monitoring cardiac mechanical activities.

SCG signal morphology is affected by different factors, such as breathing phase, patient posture, sensor position, etc. Machine learning algorithms might be used to group SCG events into groups where each group contains similar events.

This study was supported by NIH R44HL099053.

This grouping improves the quality of the signal ensemble averages, which allows more accurate estimation of SCG features. For example, it was suggested [8] that inspiratory and expiratory SCG events have different morphologies. Later studies [9]–[12] revealed that SCG signal morphology depended more on the lung volume rather than the direction of respiratory flow (i.e., inspiration vs. expiration).

This study proposes and tests a new feature extraction method that adaptively creates a feature vector. The proposed feature extraction algorithm was then used to classify SCG events into low and high lung volume (LLV and HLV) classes. The performance of the proposed method was compared to the available non-adaptive feature extraction methods [13]. These feature extraction algorithms are described in section II. Section III includes the instrumentations and experimental setup. Results are presented and discussed in sections IV and V, respectively, followed by conclusions in section VI.

## II. Feature Extraction Algorithm

The time-domain feature extraction method for SCG signals proposed by Zakeri [13] divides each SCG event into a fixed number of equal-width bins. It then uses the signal average within each bin as a feature. The grouped averages from the bins constitutes the feature vector, which will have a reduced dimensionality compared with the original signal. However, in this method, there may be bins that lack additional useful information (i.e., may have redundant values) and, hence, do not significantly improve the classification accuracy.

The feature extraction algorithm proposed in this paper aims at solving this potential issue by removing possible redundancies. In addition to dimensionality reduction, the proposed algorithm prioritizes the parts of the SCG signal that has more variations by assigning more bins to those parts.

The algorithm begins by specifying a threshold, which is a certain fraction of the signal's peak-to-peak amplitude. The threshold is chosen such that it, later, would result in a certain number of bins. In the next step the entire SCG event is considered as one bin and the signal standard deviation within the bin is calculated. When the standard deviation is above the specified threshold, the bin is divided into 2. For the new bins to approximately have equal lengths, it is preferable (but not necessary) that the SCG event contains power of 2 data points.

The standard deviation calculations are then repeated for the newly created bins, and more new bins will be created until every bin meets the threshold criteria. The outputs of this process are two vectors, where one vector contains the adaptively-spaced bins and the other contains the signal averages within each bin. The pseudo-code for the proposed algorithm is provided in Algorithm 1.

## III. HUMAN STUDIES

### A. Participants

The experimental protocol used in this study was approved by the institutional review board of the University of Central Florida, Orlando, FL. A total of 7 adult males participated in our study. The subjects provided their informed consent and reported no history of cardiovascular disease verbally. The age, height, and weight of the subjects were obtained and are reported in Table I.

### B. Experiments

For breathing pattern and tidal volume consistency, the subjects were instructed on how to breathe. For this purpose, a ventilator (Model: 613, Harvard Apparatus, South Natick, MA) was used to train subjects to breath with the same respiratory rate and inspiratory:expiratory (I:E) ratio. The respiratory rate and I:E ratio were set to 12 breath per minute and 1:3, respectively. The tidal volume (TV) for each subject was calculated in real time as time integral of the respiratory flow rate signal. During the experiment, TV was displayed on a computer screen and was about 10 to 15 mL/kg for all subjects. The subjects rested on a folding bed with their chest tilted at 45 degrees and the signals of interest were recorded for 2 trials of 5 minutes each.

### C. Instrumentation

A triaxial accelerometer (Model: 356A32, PCB Piezotronics, Depew, NY) measured all SCG signals. A signal conditioner (Model: 482C, PCB Piezotronics, Depew, NY) with a gain factor of 100 was used to amplify the accelerometer output. The accelerometer was attached with a double-sided medical-grade tape at the 4$^{th}$ intercostal space and the left sternal border. The sensor location was chosen to attain a high signal-to-noise ratio [14], [15]. The accelerometer z-axis was perpendicular to the chest surface of the subject, while the y- and x-axes were aligned parallel to the mediolateral and axial directions, respectively. The respiratory flow rate was measured using a spirometer (Model: A-FH-300, iWorx Systems, Inc., Dover, NH), that was calibrated by the manufacturer. The expiration and inspiration produced negative and positive flow rate signal amplitudes, respectively. A Control Module (Model: IX-TA-220, iWorx Systems, Inc., Dover, NH) was used to simultaneously acquire the voltage signal for respiratory flow rate, ECG, and SCG signals.

All above signals were simultaneously acquired with sampling rate of 10 kHz. To remove the remaining respiratory sound noise, the SCG signals were filtered using a low-pass filter with a cut-off of 100 Hz since lung sounds have significant energy above this cut-off frequency [16]. All signals were processed using Matlab (R2015b, The MathWorks, Inc., Natick, MA).

### D. SCG Event Pre-processing

The SCG events in each signal were found using matched filtering with a template consisting of a previously identified SCG. The lung volume signal was then used to group SCG events into two groups of high and low lung volume. The SCG events that occurred during high and low lung volumes were called HLV and LLV SCG events, respectively (Fig. 1). The SCG events duration were chosen to contain 4096 (corresponding to ~ 700 milliseconds), which was sufficiently long to contain the SCG event. This was to ensure that all signals had equal length, and that the number of points for each event equals a power of 2 to best satisfy the algorithm's requirement.

### E. SCG Feature Extraction and Classification

Feature extraction included dividing the SCG signal into bins (16, 32, 64, 128, 256, 512, and 1024), calculating the mean of each bin, and treating this as the feature vector. This corresponds to feature vectors of length 16, 32, 64, 128, 256, 512, and 1024. This method was adopted from Zakeri [13]. The original method divided the signal into evenly spaced bins, and the new proposed method divides the signal into bins adaptively

---

**Algorithm 1** Adaptive-width binning of SCG signals

1: **Input:** SCG events
2: **Output:** *Indices* for adaptive feature vector
3:     $Y$ = SCG event, $\exists x \in Z^+$ such that length($Y$) = $2^x$
4:     $T$ = Threshold
5:     *Indices* = [0, length($Y$) − 1]
6:     $a \leftarrow 0$, $b \leftarrow 1$
7:     $T \leftarrow \alpha \times$ (max($Y$) − min($Y$)), where $\alpha \in [0, 1]$
8:     **while** *Indices*[$a$] ≠ length($Y$) − 1 **then**
9:        **if** STD($Y$[*Indices*[$a$]:*Indices*[$b$]]) > $T$ **then**
10:           $C \leftarrow$ floor((*Indices*[$b$] − *Indices*[$a$]) / 2)
11:           *Indices* $\leftarrow$ sort([*Indices*, $C$])
12:        **else**
13:           $a \leftarrow a + 1$
14:           $b \leftarrow b + 1$
15:        **end if**
16:    **end while**

TABLE I    OVERVIEW OF THE SUBJECTS' CHARACTERISTICS (MEAN ± SD).

| | |
|---|---|
| Age (years) | 29.4 ± 4.5 |
| Height (cm) | 173.1 ± 9.8 |
| Weight (kg) | 82.2 ± 18.3 |
| Number of subjects | 7 |

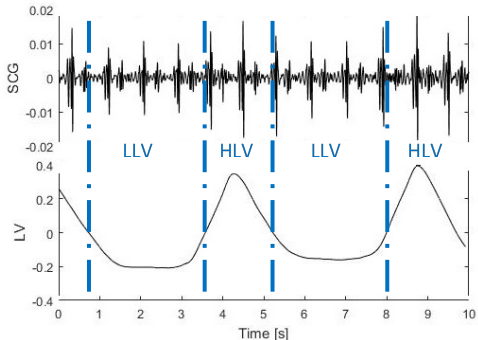

Fig. 1. (top) A 10-sec portion of the SCG signal, (bottom) Lung volume signal that was used to group SCG events into two groups of high and low

spaced using the introduced algorithm. However, when the proposed method is used, the location of the bins may be different for each SCG event, resulting in the possible transformation of each event to a different feature space. Hence, the adaptive method was instead applied to the ensemble average of all SCG events, and single set of bins was determined. The resulting bins were then used to create the feature vectors for the individual SCG events. After bins were finalized, the signal variability in each bin for the individual SCG events were checked to ensure that the variability does not exceed the threshold by more than 5%.

A support vector machine (SVM) algorithm was used to classify the SCG events into two classes of LLV and HLV. The classifier used was the Radial Basis Function (RBF) SVM. The subject-specific training scenario was employed such that a different classification model was built for each individual subject. There was a total of 3986 samples (i.e. SCG events), 1813 HLV and 2173 LLV. Per subject, there was an average of 259 ± 48 HLV, 310 ± 42 LLV, and 569 ± 76 total samples.

The $k$-fold cross-validation method ($k$=10) was used to evaluate the identification accuracy [17]. The accuracy was defined as the ratio of the number of correctly identified samples to the total number of samples in the test set. For each subject, the final accuracy of the model was obtained by averaging the 10 accuracies resulted from the $k$-fold cross-validation. The $F_1$ score was used as another evaluation metric [18]. The $F_1$ score was calculated as the harmonic mean of *sensitivity* and *precision*,

$$F_1 = 2 \times (sensitivity \times precision) / (sensitivity + precision) \quad (1)$$

where *sensitivity* and *precision* were defined as,

$$sensitivity = TP / (TP + FN) \quad (2)$$

$$precision = TP / (TP + FP) \quad (3)$$

where *TP*, *FN*, and *FP* were true positive, false negative, and false positive, respectively. The hyper parameters of the SVM models were chosen through an exhaustive grid search, with the goal of maximizing the 10-fold cross-validation accuracy. The signal processing and machine learning steps for classification of LLV and HLV SCG events are shown in Fig. 2. The feature extraction and machine learning analysis was implemented with Python libraries Scikit-Learn [19] and SciPy/NumPy [20].

IV. RESULTS

Fig. 3 shows the proposed adaptive algorithm (adaptively-spaced bins) compared to the non-adaptive method (evenly-spaced bins) used in [13] for 16 bins.

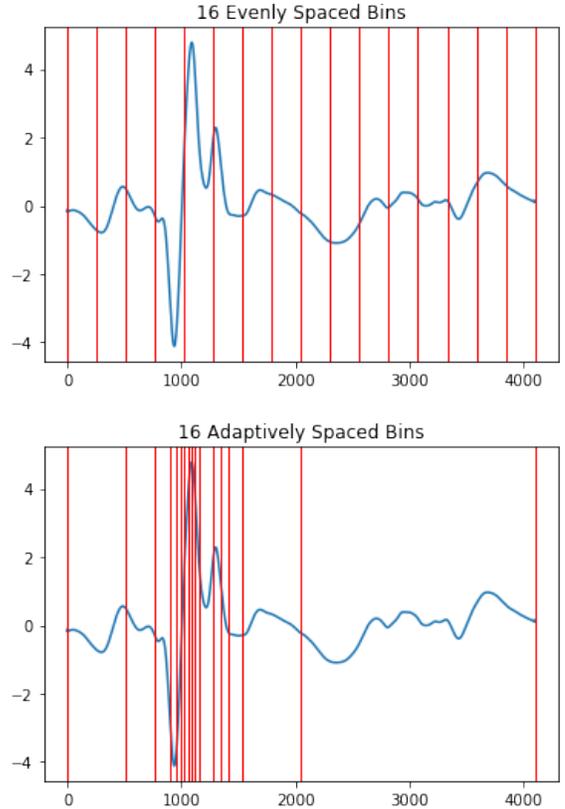

Fig. 3. (top) Equal-width bins configuration adopted from [13] vs (bottom) adaptive-width bins configuration proposed in this study for a feature vector length (i.e., number of bins) of 16.

TABLE II  AVERAGE ACCURACIES AND $F_1$ SCORES OF SVM MODELS FOR DIFFERENT NUMBER OF BINS WHEN USING EQUAL-WIDTH (EW) AND ADAPTIVE-WIDTH (AW) BINS.

| Bins | Acc. EW | Acc. AW | $F_1$ EW | $F_1$ AW |
|---|---|---|---|---|
| 16 | 0.65 ± 0.07 | 0.90 ± 0.05 | 0.63 ± 0.08 | 0.91 ± 0.05 |
| 32 | 0.66 ± 0.07 | 0.91 ± 0.04 | 0.60 ± 0.14 | 0.92 ± 0.04 |
| 64 | 0.67 ± 0.07 | 0.91 ± 0.04 | 0.65 ± 0.06 | 0.93 ± 0.04 |
| 128 | 0.70 ± 0.06 | 0.92 ± 0.04 | 0.70 ± 0.08 | 0.92 ± 0.04 |
| 256 | 0.83 ± 0.07 | 0.92 ± 0.04 | 0.82 ± 0.08 | 0.92 ± 0.04 |
| 512 | 0.87 ± 0.07 | 0.92 ± 0.04 | 0.87 ± 0.07 | 0.91 ± 0.05 |
| 1024 | 0.91 ± 0.05 | 0.92 ± 0.04 | 0.89 ± 0.07 | 0.91 ± 0.05 |

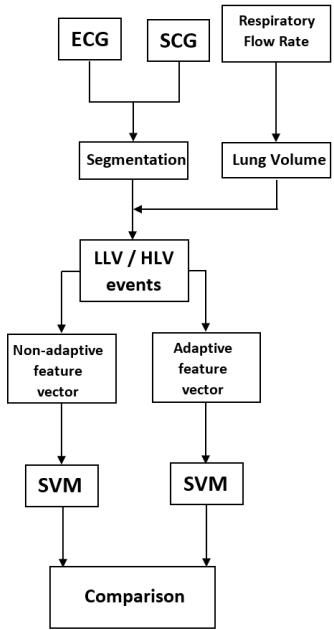

Fig. 2. Block diagram describing the signal processing and machine learning steps used in the current study.

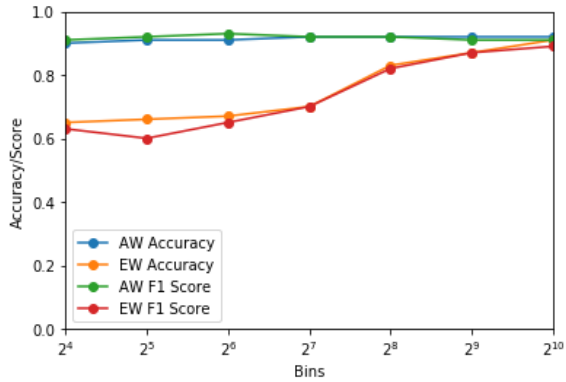

Fig. 4. Performance trend for SVM models using both AW and EW bins as the number of bins increase. The proposed adaptive feature extraction algorithm resulted in a higher accuracy and $F_1$ score consistently.

The classification of SCG events using the equal-width (EW) and adaptive-width (AW) bins began with a relatively small number of bins (16 bins). The performance analysis of the models was first done with these wider bins since it was less computationally expensive. The number of bins were then increased to find the number of bins at which the accuracy of the EW model reaches a plateau. This process is similar to mesh independency assessment in finite element analysis [21]–[30]. The identification accuracy for the SVM model for EW and AW bins feature vectors are listed in Table II for 7 different number of bins of 16, 32, 64, 128, 256, 512, and 1024. As the number of bins increased from 16 to 1024, the accuracy of the EW increased from 65% to 91% while the accuracy of the AW seemed to be stay around 91% for all number of bins used. Similarly, the $F_1$ score of the model using EW raised from 63% to 89% by increasing the number of bins from 16 to 1024. However, the $F_1$ scores of the model using adaptive feature vector were similar for different bin numbers. The accuracies and $F_1$ scores for the models using both the AW and EW bins converged as the number of bins increased, as displayed in Fig. 4. The accuracy and $F_1$ scores reached their maximum at a much smaller number of bins for the AW method, which can save calculation cost.

## V. DISCUSSIONS

### A. Algorithm Performance

The performance of the SVM using AW or EW bins was comparable for high number of bins (1024 in the current study). The performance remained relatively constant from 16 to 1024 bins for AW bins, whereas reducing the number of bins significantly worsened the classification performance when EW bins were used. These results demonstrated that the AW bins resulted in higher accuracies at smaller number of bins, and therefore it was more computationally efficient. The improved performance of the adaptive feature extraction method might be due to higher density of bins around SCG1 (the relatively strong signal region around index 1000 of the SCG event shown in Fig. 3). SCG1 contains important physiological events such as aortic valve opening, mitral valve closure, isovolumic contraction, and rapid ejection [31], [32]. Therefore, focusing on this segment of the SCG event might increase the SVM classification accuracy.

### B. Limitations

The primary limitation of the study is the small number of subjects and SCG event samples. More SCG events might result in a higher classification accuracy of the SVM models using both the AW and EW bins. Therefore, future studies need to enroll a larger number of subjects.

## VI. CONCLUSIONS

In this study, a SVM machine learning method was developed to identify the SCG events occurring during different phases of lung volume. To select features, an adaptive feature extraction algorithm was proposed. The proposed algorithm was found to be an efficient, reliable, and accurate approach to extract SCG features compared to the available feature extraction methods in literature. The feature extraction algorithm proposed in this study can be used in the analysis of other biomedical signals. In this study, the signal variability in temporal bins was measured by calculating the signal standard deviation. The proposed algorithm performance might be improved by employing other indicators of the signal complexity such as spectral entropy and average of the absolute values of the signal gradient (to measure the variation in slope).